\documentclass[aps,prb,twocolumn,showpacs,amsmath,floatfix]{revtex4}

\usepackage{graphicx}

\begin{document}

\draft

\title{Theory of hypothetical ferroelectric superlattices
incorporating head-to-head and tail-to-tail 180$^\circ$ domain
walls}

\author{Xifan Wu and David Vanderbilt}

\address{Department of Physics and Astronomy, Rutgers University,
Piscataway, New Jersey 08854-8019, USA}

\date{\today}

\begin{abstract}
While electrical compatibility constraints normally prevent
head-to-head (HH) and tail-to-tail (TT) domain walls
from forming in ferroelectric materials, we propose that such
domain walls could be stabilized by intentional growth of atomic
layers in which the cations are substituted from a neighboring
column of the periodic table.  In particular, we carry out
predictive first-principles calculations of superlattices in which Sc,
Nb, or other substitutional layers are inserted periodically into
PbTiO$_3$.  We confirm that this gives rise to a domain structure
with the longitudinal component of the polarization alternating from
domain to domain, and with the substitutional layers serving as HH
and TT domain walls.  We also find that a substantial transverse
component of the polarization can also be present.
\end{abstract}


\maketitle

\narrowtext

\marginparwidth 2.7in
\marginparsep 0.5in

\def\dvm#1{\marginpar{\small DV: #1}}
\def\xwm#1{\marginpar{\small XW: #1}}

%\def\dvm#1{}  %  Null definition
%\def\xwm#1{}  %  Null definition

%%%%%%%%%%%%%%%%%%%%%%%%%%%%%%%%%%%%%%%%%%%%%%%%%%%%%%%%%%%%

In recent years, first-principles calculations have been used
very successfully to explore new superlattice structures based on
ABO$_3$ perovskites and to predict their physical properties.
\cite{Sai,Neaton,Johnston,George}  Because of the vast number of
combinations that can be assembled from the elemental building
blocks, first-principles calculations can provide a useful way to
screen for desired physical properties and to study
interesting possibilities theoretically before they are
grown experimentally.\cite{Lee, Rijnders} For example, Sai {\it et
al.}\ predicted that a compositional perturbation that breaks
inversion symmetry could increase the piezoelectric response,
\cite{Sai} and an enhancement of the spontaneous polarization
in BaTiO$_3$/SrTiO$_3$ superlattices was predicted by
Neaton {\it et al.}\ under certain conditions of
epitaxial strain.\cite{Neaton} Also, the breaking of tetragonal symmetry
has been predicted for
certain BaTiO$_3$/SrTiO$_3$ superlattices,\cite{Johnston}
and new phase diagrams have been predicted
for superlattices compose of Pb(Sc$_{0.5+\mu}$Nb$_{0.5-\mu}$)O$_3$
with certain layer sequences of composition $\mu$.\cite{George}

In the present work, we propose a new kind of superlattice structure
which might be grown artificially via controlled layer-by-layer
epitaxy techniques such as molecular-beam epitaxy (MBE) \cite{MBE}
or pulsed laser deposition (PLD).\cite{PLD}
These contain 180$^\circ$ ferroelectric domain
walls, but of an unusual kind.  Normally, 180$^\circ$ domain
walls in a tetragonal ferroelectric such as PbTiO$_3$
form so as to obey electrical compatibility,
leading to 180$^\circ$ domain walls running parallel
to the tetragonal (polarization) axis.
This occurs because a tilting of the 180$^\circ$ domain wall would lead
to a polarization-induced bound charge on the domain wall,
leading to a large and unfavorable electrostatic energy.  Forcing
the introduction of a ``head-to-head`` (HH) or a ``tail-to-tail''
(TT) domain wall perpendicular to the polarization direction
would presumably cause it to become strongly metallic due to neutralization
by free carriers.

\begin{figure}[b]
\includegraphics[width=3.3in]{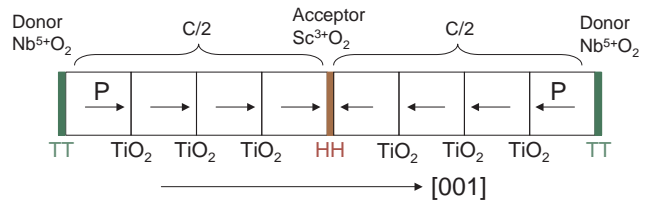}
\caption{\label{fig1} Sketch of PbTiO$_3$-based ferroelectric 
superlattice in which head-to-head and tail-to-tail 180$^\circ$
domain walls coincide with acceptor (Sc) and donor (Nb)
substitutional layers respectively.}
\end{figure}

However, we propose that intentional insertion of chemical
substitutions in certain atomic layers during growth (``delta
doping'') could provide an approximate cancellation of this bound
charge, allowing for the fabrication of new kinds of superlattice
structures containing 180$^\circ$ HH and TT domain walls, as
illustrated in Fig.~\ref{fig1}.  In this PbTiO$_3$ example, the
compensation is realized by substituting certain layers of Ti atoms
by donor or acceptor atoms drawn from neighboring columns of the
Periodic Table.  If the substitutional charge density can be chosen
to provide an approximate cancellation of the bound polarization
charge of the HH or TT domain wall, then one may hope
to arrive at a configuration that is both neutral and energetically
favorable.

In this paper, we carry out first-principles density-functional
calculations of superlattice structures such as the 8-cell one
shown in Fig.~\ref{fig1} as a ``proof of concept'' for this
idea.  We find that this arrangement, with HH and TT domain walls at
the acceptor and donor layers, respectively, is a robust
configuration in which there is no gap closure (i.e., the
superlattice remains insulating).  Somewhat surprisingly,
we also find that a small transverse component of the polarization
develops, leading to an overall spontaneous polarization of the
supercell in the [100] direction.  This can be understood as
arising from the imperfect compensation in this case.

We first select a host material for our superlattice, choosing
a tetragonal material and letting its
$c$ axis be in the superlattice growth direction.  In order
to attain perfect compensation by a donor or acceptor layer of
areal charge density $\pm e/a^2$, where $a$ is the
in-plane lattice constant, we would need a material whose spontaneous
polarization $P_{\rm s}$ is $\pm e/2a^2$.  Roughly,
$a\simeq4$\,\AA\ for most perovskites, in which case
the matching condition is $P_{\rm s}=0.50\,$C/m$^2$.  Taking
KNbO$_3$, BaTiO$_3$, and PbTiO$_3$ in their tetragonal
phases, these have $P_{\rm s}=0.40\,$C/m$^2$, $0.27\,$C/m$^2$,
and $0.75\,$C/m$^2$, respectively.  None of them is ideal, but
we have chosen to carry out our pilot calculations here using PbTiO$_3$,
whose polarization is somewhat larger than the nominal matching value.
PbTiO$_3$ is a good test system, being a well-studied
prototypical ferroelectric material having a simple phase behavior,
remaining in the tetragonal ferroelectric phase from $T=0$ up to well
above room temperature.  Experimentally, alloy host materials such as
Pb$_{1-x}$Sr$_x$TiO$_3$ or PbZr$_{1-x}$Ti$_x$O$_3$ might be used
to tune the concentration to achieve more ideal matching of the
polarization, but we have decided to limit ourselves to pure
compounds for this first study.

We construct a 40-atom supercell by arranging 8 unit cells of bulk
tetragonal PbTiO$_3$ in an $a\times a\times 8c$ supercell, with
initial atomic coordinates chosen to be those of the
relaxed bulk ferroelectric structure with polarization pointing
in the [001] direction throughout the supercell.
We then substitute the Ti atoms in the first TiO$_2$ layer by
donor atoms, and the Ti atoms in the fifth TiO$_2$ layer
by acceptor atoms, as shown in Fig.~\ref{fig1}.  
We choose Nb (5+) as the donor and Sc (3+) as the acceptor
because these atoms are commonly found on the
B site of other perovskite ABO$_3$ oxides.
We assume perfect control of the
layer-by-layer composition, resulting in an ideal
PbNbO$_{3}$/3$\times$PbTiO$_{3}$/PbScO$_{3}$/3$\times$PbTiO$_{3}$
structural sequence as shown Fig.~\ref{fig1}.

We then relax the structure under a symmetry constraint that preserves
the tetragonal $P4mm$ symmetry (i.e., atom displacements are allowed
only in the $z$ direction).  The lattice constant is fixed in the
in-plane directions, corresponding to conditions of epitaxial growth
on a bulk tetragonal PbTiO$_3$ substrate.  We also fixed the lattice
constant along the [001] direction, assuming that the volume relaxations
resulting from the compensation would be small.  All calculations
were performed at zero temperature.

The first-principles calculations were carried out using the Vienna
{\it ab-initio} simulations package (VASP)\cite{VASP}, in which
density-functional theory within the local-density approximation \cite{LDA}
is implemented.
A plane-wave basis set and projector augmented-wave
potentials\cite{PAW} were used with a 400 eV energy cutoff for the
electronic wavefunctions. Brillouin zone
integrations were performed on a Monkhorst-Pack mesh of dimensions
4$\times$4$\times$2 and 10$\times$10$\times$2 for structural relaxation
and density-of-states (DOS) calculations respectively.  The ionic
relaxation was considered to be achieved when the Hellmann-Feynman
forces on the ions were less than 3\,meV/\AA.

We find that the structural relaxation converges towards a
structure in which the polarization reverses in the top half of the
unit cell and the system acquires
an $M_z$ mirror symmetry across the planes containing the
donor and acceptor atoms, converting these layers into
180$^\circ$ TT and HH domain walls as anticipated in
Fig.~\ref{fig1}.  When we observed this apparent convergence towards
the higher-symmetry $P4/mmm$ structure, we carried out the final
relaxation with this $P4/mmm$ symmetry enforced, and confirmed that the
resulting structure had a lower energy.

In order to test whether this $P4/mmm$ is really the structural
ground state, we gave additional small displacements to the atoms
in order to break the tetragonal symmetry, and carried out further
relaxations.  Surprisingly, we found that the system
is unstable to the growth of an in-plane component of the polarization.
We tested both initial [100] displacements and initial [110]
displacements (preserving, in both cases, the $M_z$ symmetry),
giving rise to $Pmm2$ and $Amm2$ structures, respectively.
We find that the relaxed $Pmm2$ structure has an energy
0.065\,eV lower than that of the $P4/mmm$ structure, while the
relaxed $Amm2$ one is 0.057\,eV lower than the $P4/mmm$.
Thus, we find that the $Pmm2$ structure is our final structural ground state.

\begin{figure}
\includegraphics[width=2.7in]{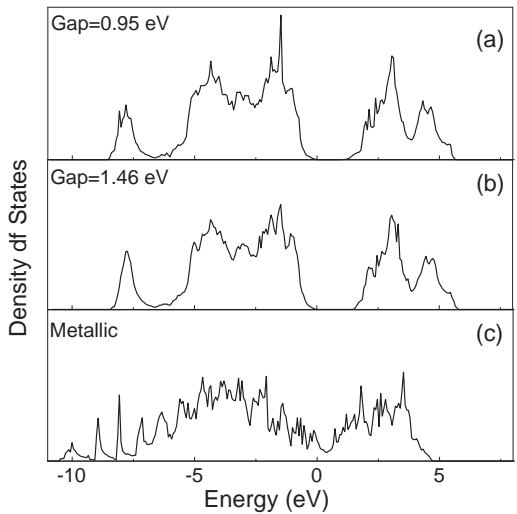}
\caption{\label{fig2}
Density of states (DOS) of 40-atom PbTiO$_3$ supercells considered
in the text.  (a) $P4/mmm$ supercell of Fig.~\protect{\ref{fig1}}
containing HH and TT domain walls compensated by impurity layers.
(b) Same, but with in-plane component of polarization ($Pmm2$
symmetry).  (c) Reference structure in which polarization reverses
as in Fig.~\protect{\ref{fig1}}, but no compensating impurity layers
are present.}
\end{figure}

To get a better sense of the electronic structure of the
relaxed structures, we computed the electronic density of
states (DOS).  The results, shown in Fig.~\ref{fig2}(a) and (b) for the
$P4/mmm$ and $Pmm2$ structures respectively, indicate that
the compensation is successful and the entire superlattice is
insulating.  (The DOS for the $Amm2$ structure looks
very similar to these.)  The final gap of 1.46\,eV
is moderately smaller, but not drastically smaller, than
the gap of 1.78\,eV for bulk tetragonal PbTiO$_3$.
(As these gaps are computed using the LDA,
they should be regarded as underestimates of the
true gaps.)

We also checked what happens
if there is no compensation between the polarization-induced bound
charge and the donor/acceptor layer.  To do this, we constructed
a 40-atom $P4/mmm$ supercell similar to that of Fig.~\ref{fig2}(a),
but without introducing the impurity layers.  (Actually, this was
done by imposing the displacement pattern of the
ferroelectric tetragonal ground state of PbTiO$_3$, but with
opposite signs in the bottom and top halves of the supercell.)
The result is shown if Fig.~\ref{fig2}(c), where it is clear that
the structure is strongly metallic.  Thus, the compensation by the
donor and acceptor layers is essential in order to arrive at an
overall insulating structure.

We return now to the pattern of ground-state atomic displacements that
we calculated for the $P4/mmm$ and $P2mm$ structures of
Fig.~\ref{fig2}(a) and (b), and interpret these in terms of a local
polarization in each unit cell.  To do this, we model the
local polarization as\cite{Meyer}
\begin{equation}
{\bf P}^{(i)} = {\frac{e}{\Omega_{c}}}\sum_{\alpha}w_{\alpha}
{\bf Z}^{\ast}_{\alpha}\cdot {\bf u}^{(i)}_{\alpha} \;,
\label {eq:P_local}
\end{equation}
where $i$ is the Pb-centered unit cell index, $\alpha$ runs over all
atoms, and $w_{\alpha}$ is a weight factor that is
unity inside the cell, zero outside, and takes appropriate fractional
values for atoms shared with a neighboring cell.\cite{Meyer}
For the dynamical effective charge tensors ${\bf Z}^{\ast}_{\alpha}$,
we use bulk PbTiO$_3$ values of
$Z^{\ast}$(Pb)=3.90, $Z^{\ast}$(Ti)=7.06,
$Z^{\ast}_{\parallel}$(O)=$-$5.83, and $Z^{\ast}_{\perp}$(O)=$-$2.56
following Ghosez et al.\cite{Ghosez}  We also arbitrarily adopt
$Z^{\ast}$(Nb)=8.06 and $Z^{\ast}$(Sc)=6.06 based on the rough
expectation that these atoms should resemble Ti except a shift
by $\pm e$ of the core charge.

\begin{figure}
\includegraphics[width=3.1in]{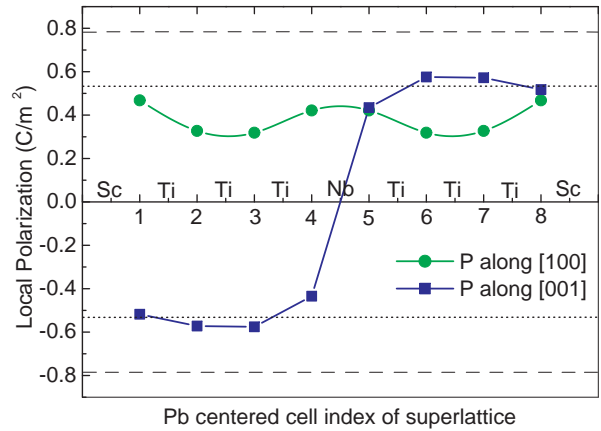}
\caption{\label{fig3}
Local polarization profile of the relaxed $Pmm2$ structure.  The
[001] component changes sign at the dopant layers, while the
[100] component is roughly uniform throughout the supercell,
giving rise to an overall polarization in the [100] direction.
Dotted lines indicate the ``ideal'' value 0.5e/a$^2$;
dashed lines represent the spontaneous polarization of bulk tetragonal
PbTiO$_3$.}
\end{figure}

The results of this analysis are shown in Fig.~\ref{fig3}, which shows
the local polarizations in the ground-state $Pmm2$ structure of the
superlattice.  The local polarization in the [001] direction changes
sign across the dopant layers, confirming that HH and TT domain walls
are formed.  In addition, there is a roughly uniform polarization
in the [100] direction, although it is slightly enhanced in the
vicinity of the HH and TT domain walls where the [001] polarization
is crossing through zero.  Thus, the entire superlattice structure
is antiferroelectric as concerns the [001] polarizations, but is
ferroelectric in the [100] direction.  (The $P_z$ profile of the
$P4/mmm$-constrained structure is not shown, but it is almost
identical to that of the $Pmm2$ structure.)

In order to understand our results, recall that PbTiO$_3$ has a
spontaneous polarization of $0.734e/a^2$, somewhat larger than the ideal
value of $0.5e/a^2$ that would give perfect compensation of the donor and
acceptor layers.  If each domain were to develop its
full bulk spontaneous polarization, there
would be substantial charges ($\pm0.47e/a^2$) at the donor layers due
to overcompensation by the domain walls. The resulting
depolarizing field would reduce $P_z$ to a value much closer
to, but still larger than, the reference value of $0.5e/a^2$.  This
is precisely what we see in Fig.~\ref{fig3};
in the middle of each domain, $P_z$ attains a value just slightly larger
than $0.5e/a^2$.

The appearance of the polarization in the in-plane direction can be
understood based on the concept of polarization rotation.  Many previous
works \cite{rotation}
have provided support for a picture in which the crystal is regarded
as having a preferred {\it magnitude} of the polarization, with a rather
weak crystalline anisotropy determining its {\it direction}.
If the polarization is prevented from
attaining its preferred bulk value in the [001] direction, it can
still remain close to this preferred magnitude by tilting off this
axis.  Indeed, the resulting magnitude of the polarization in the middle
layers of the slab, $0.57e/a^2$,
is only slightly lower than the preferred bulk value.  Moreover, this
picture naturally helps explain why the in-plane component of the
polarization grows near the HH and TT domain walls, where $P_z$
is in the process of crossing through zero.

\begin{figure}
\includegraphics[width=2.6in]{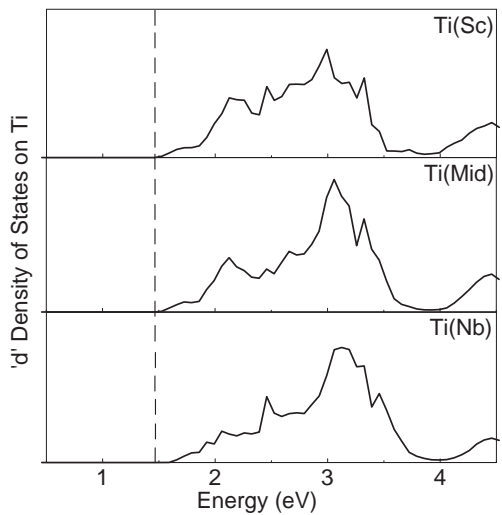}
\caption{\label{fig4} DOS projected onto three different Ti atoms
in the $Pmm2$ supercell.  Ti(Sc) denotes the Ti atom in the TiO$_2$
layer closest to the ScO$_2$ layer; Ti(Nb) denotes the Ti atom in
the TiO$_2$ layer closest to the NbO$_2$ layer; and Ti(Mid) denotes
a Ti atom in the middle of the supercell, between the two dopant
layers.}
\end{figure}

Fig.~\ref{fig4} gives further evidence of the physical picture
outlined above.  Here, we have plotted the site-projected
DOS for Ti atoms sitting on three different TiO$_2$
layers of the $Pmm2$ supercell.  Specifically, Ti(Nb) is a
Ti atom sitting next to the donor-compensated (NB) plane; Ti(Sc) sits
next to the acceptor-compensated (Sc) plane; and Ti(Mid)
is a Ti atom midway between the two impurity layers.
The zero of energy is at the valence-band maximum.
The vertical dashed line provides a reference that helps clarify
that the spectral weight of the conduction band, and in
particular its low-energy edge, shifts to higher energies
as one progresses from the Ti(Sc) to the Ti(Nb) layer.  The smallness
of these shifts confirms that the compensation is quite good, and
the small shift that does occur can be attributed to the
leftover depolarization field in the [001] direction.  Furthermore,
the sign of this effect is consistent with our previous
conjecture that the domain walls are slightly under-compensated.

It is interesting to ask what will happen for
superlattices of longer and longer period in the $c$ direction,
for a given material system.  Unless the
compensation is perfect, the depolarizing fields resulting from
the imperfect compensation should result in an electrostatic
potential having a sawtooth dependence on $z$.  This
should be sufficient to close the gap, and drive the
superlattice structure metallic, when the
amplitude of this sawtooth potential exceeds the band gap.
We can estimate the critical superlattice period for the
PbTiO$_3$/Sc/Nb system by recalling that we estimated
a net bare charge of $2 P_s - e/a^2\simeq 0.47\,e/a^2$
or about $0.51$\,C/m$^2$.  Taking the theoretical value of the bulk
band gap of $E_g=1.8$\,eV and a bulk dielectric constant of about 110,
we estimate that the depolarization field would close the band gap
when the distance $L$ between impurity layers exceeds a critical
value of about $75$\,\AA\ or about 19 PbTiO$_3$ layers.

If it is desired to increase the superlattice period and still keep
an insulating structure, then the precision of the compensation
should be improved.  Two approaches suggest themselves.  First,
one could make use of mixed dopant layers of composition Ti$_{1-x}$Sc$_x$ or
Ti$_{1-x}$Nb$_x$, with $x$ tuned to obtain
the correct degree of compensation for a given bulk material.
Second, the bulk material itself may
be varied in order to adjust its spontaneous polarization closer
to the ideal value of 0.5\,$e/a^2$.  For example, replacing
PbTiO$_3$ by PbZr$_x$Ti$_{1-x}$O$_3$ (PZT) with $x \simeq$ 0.2
may accomplish this.\cite{pzt}  Using such strategies, it
may be possible to construct long-period insulating superlattices,
perhaps even with periods large enough to be of use for optical
grating applications.

In summary, we have calculated the properties of a particular
realization of a hypothetical class of ferroelectric superlattice
structures containing head-to-head and tail-to-tail 180$^\circ$
domain walls compensated and stabilized by the intentional
insertion of atomic charged-impurity layers.  In the system we
chose for study, in which Sc and Nb substitutional layers provide the
compensation in PbTiO$_3$, the spontaneous polarization of the
bulk material is larger than the ideal value; consequently,
the superlattice develops an in-plane ferroelectric polarization,
while also developing the expected antiferroelectric domain structure
in the growth direction.  This appears to be a novel way of forcing
polarization rotation into the in-plane orientation in PbTiO$_3$,
and it would be interesting to explore how the superlattice properties
(spontaneous polarization, Curie temperature, etc.) could be tuned
by the growth conditions (e.g., superlattice period).  More
generally, it would be interesting to investigate other host
perovskite ferroelectric materials, other kinds of compensating
layers (including, perhaps, dopants from column II or VI of the
periodic table), and alloy compositions of the bulk and/or
dopant layers.  With a more suitable matching of the compensation,
the driving force for polarization rotation could be reduced, and a true
$P4/mmm$ superlattice structure could presumably be stabilized.  In
fact, it also seems possible that a normally rhombohedral bulk ferroelectric
like BaTiO$_3$ whose spontaneous polarization is {\it smaller} than the
ideal value of 0.5\,$e/a^2$ could have its polarization rotated in
the opposite way, into alignment with the $z$-axis, since in this
case the compensating layers would enhance the polarization in that
direction.  Finally, much longer-period superlattices might be obtained
by very accurate matching of the compensation.  There are thus many
open avenues for future theoretical and experimental studies.

%%%%%%%%%%%%%%%%%%%%%%%%%%%%%%%%%%%%%%%%%%%%%%%%%%%%%%%%%%%%%%%%%%%%%%%

\acknowledgments

The work was supported by ONR Grants N00014-97-1-0048
and N00014-05-1-0054
and by the Center for Piezoelectrics by Design under ONR
Grant N00014-01-1-0365.  We wish to thank M.~Cohen, 
D.~Hamann and K.~Rabe for valuable discussions.

%%%%%%%%%%%%%%%%%%%%%%%%%%%%%%%%%%%%%%%%%%%%%%%%%%%%%%%%%%

%%%%%%%%%%%%%%%%%%%%%%%%%%%%%%%%%%%%%%%%%%%%%%%%%%%%%%%%%%%


\begin{thebibliography}{0}

%%%%%%%%%%%%%%%%%%%%%%%%%%%%%%%%%%%%%%%%%%%%%%%%%%%%%%%%%%

\bibitem{Sai}
N. Sai, B. Meyer, and D. Vanderbilt, Phys. Rev. Lett. {\bf 84}, 5636 (2000).

\bibitem{Neaton}
J. B. Neaton and K. M. Rabe, Appl. Phys. Lett. {\bf 82}, 1586 (2003).

\bibitem{Johnston}
K. Johnston, X. Huang, J. B. Neaton, and K. M. Rabe, Phys. Rev. B 
{\bf 71}, R100103(2005).

\bibitem{George}
A. M. George, J. \'{I}\~{n}iguez, and L. Bellaiche, Nature {\bf 413}, 54 (2001). 

\bibitem{Lee}
H. N. Lee, H. M. Christen, M. F. Chisholm, C. M. Rouleau, and D. H. Lowndes,
Nature {\bf 433}, 395(2005).

\bibitem{Rijnders}
G. Rijnders and D. A. Blank, Nature {\bf 433}, 369(2005).

\bibitem{MBE} M. P. Warusawithana, E. V. Colla, J.N. Eckstein,
and M. B. Weissman, Phys. Rev. Lett. {\bf 90}, 036802 (2003).

\bibitem{PLD}
T. Shimuta, O. Nakagawara, T. Makino, S. Arai, H. Tabata, and T. Kawai,
J. Appl. Phys. {\bf 91}, 2290 (2002).

%\bibitem{Padilla}
%J. Padilla, W. Zhong and D. Vanderbilt, Phys. Rev. B {\bf 53}, R5969 (1996).

\bibitem{VASP}
G. Kresse and  J. Hafner, Phys. Rev. B {\bf 47}, R558 (1993);
G. Kresse and  J. Furthm\"uller, {\it ibid.}\ {\bf 54}, 11169 (1996).
See also http://cms.mpi.univie.ac.at/vasp.

\bibitem{LDA}
P. Hohenberg and W. Kohn, Phys. Rev. {\bf 136},
B864(1964); W. Kohn and L. J. Sham, {\it ibid.} {\bf 140}, A1133 (1965).

\bibitem{PAW}
P. Bl\"{o}chl, Phys. Rev. B {\bf 50}, 17953 (1994);
G. Kresse and D. Joubert, {\it ibid.} {\bf 59}, 1758 (1999).

\bibitem{Meyer}
B. Meyer and D. Vanderbilt, Phys. Rev. B {\bf 65}, 104111 (2002).

%\bibitem{He}
%L. He and D. Vanderbilt, Phys. Rev. B {\bf 68}, 134103 (2003).

%\bibitem{Lines}
%M. E. Lines and A. M. Glass, {\it Principles and Applications of 
%Ferroelectrics and Related Materials} (Clarendon Press, Oxford, 1977).

\bibitem{Ghosez}
Ph. Ghosez, J. -P. Michenaud, and X. Gonze, Phys. Rev. B, {\bf 58}, 6224(1998).

\bibitem{rotation}
H. Fu and R. E. Cohen, Nature (london) {\bf 403}, 281 (2000);
L. Bellaiche, A. Garc\'{i}a, and D. Vanderbilt {\bf 64}, R060103 (2001).

\bibitem{pzt}
J. Frantti, S. Ivanov, S. Eriksson, H. Rundl\"{o}f, V. Lantto,
J. Lappalainen, and M. Kakihana, 
Phys. Rev. B {\bf 66}, 064108 (2002).


\end{thebibliography}
\end{document}